\documentclass[referee,useAMS,usenatbib]{mn2e}
\usepackage{fourier}
\usepackage{graphicx}
\usepackage{lscape}
\usepackage{subfigure}
\title[Infall motion in G8.68-0.37]{The molecular emissions and the infall motion in the high-mass young stellar object G8.68-0.37}

\author[Ren et al.]
{{\normalsize Zhiyuan Ren$^{1}$, Yuefang Wu$^{1}$, Ming Zhu$^{2}$, Tie Liu$^{1}$,
Ruisheng Peng$^{3}$, Shengli, Qin$^{4}$, and Lixin Li$^{1,5}$} \\ \\
$^{1}${Department of Astronomy, Peking University, 100871, Beijing China, E-mail:rzy,ywu@pku.edu.cn} \\
$^{2}${National Astronomical Observatory of China, 20A Datun Road, Chaoyang District, Beijing, China}\\
$^{3}${Caltech Submillimeter Observatory}\\
$^{4}${I. Physikalisches Institut, Universit\"{a}t zu K\"{o}ln, Z\"{u}lpicher Str. 77, 50937} \\
$^{5}${The Kavli Institute for Astronomy and Astrophysics, Peking University, Yi He Yuan Lu 5, Hai Dian Qu, Beijing 100871, P. R. China}\\
}
\begin{document}
%\end{minipage}
%\date{Accepted xxxx. Received xxxx; in original form xxxx}

\pagerange{\pageref{firstpage}--\pageref{lastpage}} \pubyear{2010}

\maketitle

\label{firstpage}

\begin{abstract}
We present a multi-wavelength observational study towards the high-mass young stellar object G8.68-0.37. A single massive gas-and-dust core is observed in the (sub)millimeter continuum and molecular line emissions. We fitted the spectral energy distribution (SED) from the dust continuum emission. The best-fit SED suggests the presence of two components with temperature of $T_{\rm d}=20$ K and 120 K, respectively. The core has a total mass of up to $1.5\times10^3$ $M_{\odot}$ and bolometric luminosity of $2.3\times10^4~L_{\odot}$. Both the mass and luminosity are dominated by the cold component ($T_{\rm d}=20$ K). The molecular lines of C$^{18}$O, C$^{34}$S, DCN, and thermally excited CH$_3$OH are detected in this core. Prominent infall signatures are observed in the $^{12}$CO $(1-0)$ and $(2-1)$. We estimated an infall velocity of 0.45 km s$^{-1}$ and mass infall rate of $7\times10^{-4}~M_{\odot}$ year$^{-1}$. From the molecular lines, we have found a high DCN abundance and relative abundance ratio to HCN. The overabundant DCN may originate from a significant deuteration in the previous cold pre-protostellar phase. And the DCN should now be rapidly sublimated from the grain mantles to maintain the overabundance in the gas phase.

\end{abstract}


% 不知道infall是否各向同性
% 不同的亚毫米波段展现一样的
% accretion 有助于增加D/H，观测显示清楚。
% PMO探测到outflow说明其很延展
% HCN没有同位素可以判断光深和谱线轮廓，但其他分子基本上可以排除多成分的可能性
% new:
% 表格加上两个温度，光深。
% 450比1.3稀薄，但总质量要大一些
% ... therefore the datapoints bellow 8 micron are not considered in the SED fitting.
% 因为没有电离，所以24 micron 辐射主要来自尘埃
% infall的size scale再突出一下 (done)
% no multiple clumps... (done)
% DCN 4-3 and 3-2 计算出的参数有神马不同？
% Table 4:加上1.3mm和450 micron的峰值流量
% 在我们的观测水平上，没有发现fragmentation.
% 为了计算Dusty core的infall rate,还是采用450micron的半径
% 总的柱密度
% Using the molecular lines, we can estimate their column density $N(X)$ and abundances $f_{\rm x}$.
% recently released into the gas by the stellar emission.
% beam dilution: N(x) underestimate.
% 中心位置不再列表里，而是写在text中。(done) 18:06:23.23 -21:37:14.09
% To better learn their properties, currently we are focusing on individual massive YSOs,
% attempting to analyzing their spatial morphologies and physical parameters.
% IRAC shock emission (done)
% 表达: skewed to the blueshifted side with respect to the C18O lines.
% 流量误差由15%决定
% H13CO+ abundance

\section{Introduction}
Gravitational infall, or core collapse can take place in high mass young stellar objects (YSOs) at early stages and continue all the way to the stage of Ultra Compact (UC) H{\sc ii} regions~\citep{keto03,sol05}. As shown by theoretical works\citep[][etc.]{ji96,york02,gong09}, the infall motion is critical for initiating the high mass star formation and maintaining the accretion flow to feed the stellar mass during the subsequent evolutionary stages. However, further observations are still needed to better constrain the physical properties of the infall, including its spatial distribution, mass infall rate, chemical effect, and to understand its relation with other dynamical processes, including outflow, disk accretion, and core fragmentation. In the recent decade, extensive spectroscopic surveys~\citep[e.g.][]{wu03,fuller05,wy06,pur06,wu07} have been performed towards the potential high mass YSOs throughout the Milky Way. As a result, many infall candidates have been identified based on their spectral signatures. These sources can serve as good candidates to study the massive star birth and gas dynamics in the molecular cores. In the mean time, strong outflows are also widely detected towards those massive cores~\citep[e.g.][]{beu02,wu04,zhang07}. The infall and outflow motions should be closely related and interacting with each other throughout the star formation history.

G8.68-0.37 (G8.68 here after) is a young high-mass star forming region at a distance of 4.5 kpc~\citep{mue02}. In this region, compact multiple gas-and-dust clumps has been discovered by \citet[][L11 here after]{long11}. The dusty core is associated with strong 6.7 GHz methanol masers~\citep{walsh98}, but has no radio continuum emission, indicating that high mass stars are already formed, but have not yet ionized its surrounding gas. L11 also detected a bi-polar outflow in CO $(2-1)$. The outflow may be responsible for the shock interaction traced by the extended 4.5 $\micron$ emission (Figure 2 therein). In the mean time, the observation in HCO$^+$ $(1-0)$ suggests a plausible infall motion~\citep{pur06} which should be examined quantitatively. To improve the understanding in physical and chemical properties of this source, we performed a multi-wavelength study using both the single dish antennae and the interferometers. The next section introduces the observations and data reduction, Section 3 presents the general observational results. Section 4 describes the dust continuum and molecular line emissions, wherein the infall signature is specifically described in Section 4.3. A summary is given in Section 5.

\section{Observations and Data Reduction}
\subsection{The single dishes}
In Figure 1 we show the central positions and the beam sizes of all the observations. We have observed G8.68 using three different single-dish telescopes. In 2005, we observed HCN (3-2) and H$^{13}$CO$^+$ $(3-2)$ from the James Clerk Maxwell Telescope\footnote{JCMT is operated by the JAC, Hawaii, on behalf of the UK PPARC, the Netherlands OSR, and the Canadian NRC, see http://www.jach.hawaii.edu/JCMT/} (JCMT). The pointing center was adopted to be the coordinate of the strongest methanol maser \citep[][with a position accuracy of $1.8''$]{walsh98}, which is close to the continuum emission peak of L11 (cross in Figure 1). The mapping step is $10''$ (corresponding to 1/2 beam size), as shown in Figure 2b.

In November 2009, we observed $J=1-0$ line of $^{12}$CO, $^{13}$CO, and C$^{18}$O using the 13.7 m telescope at the Purple Mountain Observatory\footnote{http://www.dlh.pmo.cas.cn/} (PMO). The PMO observation contains a grid-mapping with a coverage of several arc minutes over the region of G8.68. In this paper, we only use the spectra at one point which is closest to the continuum peak, as shown in Figure 1. The PMO beam is much larger than the CSO and JCMT, and is significantly deviated from the continuum peak. Nevertheless, the beam has well covered the emission regions of the continuum and molecular lines. The data can thus be used to trace the gas motion on a larger scale near the core.

In May 2011, the $J=2-1$ lines of the three CO isotopologues were observed from the 10 m telescope at the Caltech Submillimeter Observatory\footnote{http://www.submm.caltech.edu/cso/} (CSO). The $^{12}$CO $(2-1)$ is observed at five symmetric points around the continuum peak with an offset of $\pm23''$ in the R.A. and Dec. directions. Their positions are shown in Figure 2a. All the single-dish spectra are discussed in detailed in Section 3.2.2.

In Table 1, we present the basic observational parameters and weather conditions for the three instruments. In Table 2, we shows the more specific observational parameters for the molecular lines. All the observations were performed in good weather conditions, with pointing accuracies better than $5''$. GILDAS software package\footnote{http://iram.fr/IRAMFR/GILDAS/} is used for the data reduction and image plot.

To measure the flux densities at different wavelengths, we also retrieved the Spitzer archival images at four IRAC bands from the GLIMPSE survey\footnote{Available at http://irsa.ipac.caltech.edu/, see also \citet{ben03}.}, and the 24 and 70 $\micron$ images from MIPSGAL\footnote{http://irsa.ipac.caltech.edu/}, and JCMT/SCUBA images at 450 and 850 $\micron$ bands, which are available at the Canadian Astronomy Data Center (CADC) repository of the SCUBA Legacy Fundamental Object Catalogue\footnote{http://www4.cadc-ccda.hia-iha.nrc-cnrc.gc.ca}.

\subsection{The Submillimeter Array}
The Submillimeter Array\footnote{The Submillimeter Array is a joint project between the Smithsonian Astrophysical Observatory and the Academia Sinica Institute of Astronomy and Astrophysics and is funded by the Smithsonian Institution and the Academia Sinica, see http://www.cfa.harvard.edu/sma/} (SMA) observations towards G8.68 are taken from the released SMA data archive. The observations are made in three epochs in the year 2007, 2008, and 2009, respectively. The observational parameters, including the calibration sources for each epoch are presented in Table 3. In all three epochs, the compact array was used, and the phase tracking center is R.A.(J2000)=18$^{\rm h}$06$^{\rm m}$23.23$^{\rm s}$, Dec.(J2000)=$-21^{\circ}37'14.19''$. The three observations have similar beam sizes for the synthesized and primary beams. In Figure 1, we only show the beams of the 2008 observation in order to have a clear appearance. The observed gas-and-dust structures (Figure 4) turn out to be smaller than the SMA primary beam. Thus the beam-edge weakening is not significant. The calibration and imaging were performed in Miriad$^{1}$. The absolute flux level has an uncertainty of $\sim15\%$. The continuum emission was subtracted from the line-free channels in each sideband. The gain solution is self-calibrated for the continuum image and then exported to the spectral line data.

We note that among the SMA data, the 2008 observation has the longest on-source integration time hence the lowest noise level. In addition, in 2008 all eight antennae of the SMA were at work, while the 2007 observation (280 GHz) only employed seven antennae. As a result, despite its higher frequency, the 2007 data has a lower angular resolution than the 2008 data (as indicated by their synthesized beam sizes, in Table 3). We therefore used the 2008 data (frequency centered at 225 GHz, or 1.3 mm) to analyze the dust continuum emission. The continuum was averaged from the line-free channels and then subtracted from the side-band spectrum. The continuum data of the two sidebands were averaged on the (u,v) plane and then converted to the image domain. After Clean and Self-calibration, the 1.3 mm continuum image has an rms noise level ($1~\sigma$) of 3.6 mJy beam$^{-1}$ (corresponding to a brightness temperature of $T_{\rm b}=0.0025$ K).

\section{Results}
\subsection{Dust continuum emission}   %section 3.1
In Figure 4, we show the continuum emissions of G8.68 from infrared to (sub)millimeter wavebands, including the IRAC 3.6, 4.5, and 8 $\micron$ emissions (RGB color image), the SMA 1.3 mm continuum emission (white contours), and the SCUBA 450 $\micron$ continuum (blue dashed contours).

Figure 4 is centered at the 1.3 mm continuum peak, the coordinates of which are R.A.(J2000)$=18^{\rm h}06^{\rm m}23.52^{\rm s}$, Decl.(2000)$=-21^{\circ}37'11''$. It is close in projection to the SMA phase tracking center (labeled with the red cross). After deconvolution with the synthesized beam, the core has an angular size of $11''\times6''$ for the 4 $\sigma$ contour ($0.24\times0.13$ pc at a distance of 4.5 kpc). It is elongated in the north-south direction ($PA=-10^{\circ}$ for the major axis), reasonably coherent with the 4.5 $\micron$ emission (green color). Since the 4.5 $\micron$ emission traces the shock interaction between the outflow and the envelope gas, it is possible that the outflow and shocks are also affecting the dust distribution, causing its observed elongation. We did not find any evidence for multiple sub-cores either in our 1.3 mm continuum or any molecular lines (Figure 5). Therefore the gas-and-dust core should have a single compact morphology, and the fragmentation is not evident on our observational scale (0.05 to 0.5 pc).

More diffused dust component can be revealed by the SCUBA 450 $\micron$ continuum emission. As shown in Figure 4, the 450 $\micron$ emission is more extended and less elongated than the 1.3 mm emission. We use the average deconvolved FWHM (full width at half maximum) radius $\langle r\rangle$ to represent the extent of the continuum and molecular line emissions. Normally $\langle r\rangle$ can be measured from the 50 \% contour level of the emission region. However, the 50 \% contour (for the continuum and molecular lines) is often close to or even smaller than beam size. We thus suggest measuring the deconvolved radius from the 10 \% contour, and adopt its 1/2 as the value of $\langle r\rangle$. For the continuum images, the 10 \% contour is not specifically plotted in Figure 4, but close to the 14 $\sigma$ and 4 $\sigma$ contour level for the 1.3 mm and 450 $\micron$ emissions, respectively. We obtained $\langle r\rangle_{\rm 450 \micron}=0.23$ pc and $\langle r\rangle_{\rm 1.3 mm}=0.08$ pc.

We also measured the integrated flux density $F(\lambda)$ of the dust core at wavelength $\lambda$. In general, we use the 4 $\sigma$ emission level as the integration area for $F(\lambda)$. As an exception, for the IRAC data, we use the region of the 4.5 $\micron$ emission (green color in Figure 4) to measure the integrated flux of all four bands, since the 4.5 $\micron$ emission has a relatively clear boundary. The 5.8 $\micron$ band shows a similar morphology with the 4.5 $\micron$, while the emissions at other two bands are much fainter and cannot be well delineated. The IRAC stellar sources in the vicinity of the core are carefully excluded from the integration area. The derived $F(\lambda)$ are shown in Table 4.

\subsection{Molecular lines}
\subsubsection{The Submillimeter Array}
Using the SMA, we have detected a number of molecular transitions of C$^{18}$O, C$^{34}$S, DCN, and CH$_3$OH. Their beam-averaged spectra towards the 1.3 mm continuum peak and their velocity-integrated intensity images are shown in Figure 5. For the CH$_3$OH, altogether we have detected 11 rotational transitions. We selected five of them with largely different $E_{\rm u}$, and presented their images in Figure 5. The physical parameters of all the molecular transitions are shown in Table 5.

As shown in Figure 5, the spectra of the molecular tracers of high-density gas mostly show lines with single peak profiles. As exceptions, there are two CH$_3$OH lines, $11_2-10_3$ ($E_{\rm u}=191$ K) and $15_{7,8}-16_{6,11}$ ($E_{\rm u}=523$ K) which show double peak profiles. However, since the remaining lines are all single-peaked, the two lines are more likely to be blended with other molecular transitions. The possible candidates for the blenders are NH$_2$CN $14_2-13_2$ (f=279.35062 GHz, $E_{\rm u}=228$ K) and HCCNC $29-28$ (f=288.07346 GHz, $E_{\rm u}=207$ K). We used two gaussian profiles to fit the blended spectra, as plotted in dotted lines in Figure 5. For each spectrum, the peak velocity of the second component is well consistent with the anticipated velocity for the blenders. With the contamination excluded, these two CH$_3$OH lines should also have single gaussian profiles.

The C$^{34}$S emission region shows an elongated morphology from the northeast to southwest ($PA=45^{\circ}$) as labeled in dashed line. The elongation agrees with the orientation of the CO outflow and 4.5 $\micron$ shock emission (Figure 4). An elongated morphology towards northeast is also shown in the low excited CH$_3$OH lines (i.e. $E_{\rm u}=33$ K and 97 K). Therefore, both the CH$_3$OH and C$^{34}$S distributions should be affected by the outflow. The C$^{18}$O $(2-1)$ also shows a non-regular morphology. However, it is biased to the south of the dust core, peaked at offset$=(0'',-2'')$, In addition, the C$^{18}$O shows a secondary clump in the southeast, peaked at offset$=(10'',-8'')$. This clump is not detected in C$^{34}$S or CH$_3$OH lines, indicating that it may be depleted in these species. The C$^{18}$O might trace cooler and less dense gas component, thus have a more extended feature than other dense-core species. For each molecular transition, we also measured the average deconvolved radius from the 10 \% contour level (and adopted its 1/2 as the value of $\langle r\rangle$). The results are shown in Table 5.

\subsubsection{The single dishes}
Figure 2 and 3 show the molecular lines detected from the single dishes. As shown in Figure 3, prominent double-peak line profiles are observed in the $^{12}$CO $(1-0)$, $(2-1)$, and also HCN $(3-2)$. For both the $^{12}$CO $(1-0)$ and $(2-1)$, the blueshifted emission peak is much stronger than the redshifted one, and the central absorption dip is well coincident with the C$^{18}$O line peak ($V_{\rm lsr}=37$ km s$^{-1}$). Such blue asymmetric $^{12}$CO lines suggest the presence of infall motion towards the core center \citep{zhou93,mard97}. For the physical explanation, when the infall occurs in the envelope which is cooler than the inner region, the gas in the front part would absorb the redshifted side of the line profile, whereas the gas in the rear part (behind the center) would increase the blueshifted emission because it is moving towards the observer. Besides the infall signature, the $^{12}$CO lines also exhibit high-velocity emission wings extending to $V_{\rm lsr}=$25 and 50 km s$^{-1}$ for the blue- and redshifted sides, respectively. This velocity range is comparable to the outflow velocities revealed by L11 (Figure 5 therein).

The two $^{13}$CO lines also have a blueshifted emission peak ($V_{\rm lsr}=35$ km s$^{-1}$) with respect to the C$^{18}$O, suggesting that the $^{13}$CO is also probably tracing the infall motion. However, because the $^{13}$CO lines are much less optically thick than the $^{12}$CO, they exhibit no central dip, but instead show an emission shoulder that continuously declines towards the redshifted side.

As shown in Figure 2a, we can see that the offset positions also exhibit self-absorbed profiles (except the southeast one). However, compared to the central spectrum, their blue- and redshifted peaks have more similar intensities. As an extreme, the southeastern spectrum have a flattened top, with the double-peak feature almost disappeared. This indicates that the infall motion (along the line of sight) should have a decline towards those offset points. And their distance from the center (0.7 pc) can therefore be taken as a lower limit for the radius of the infalling region.

%2011-9-8
As shown in Figure 3c, the HCN $(3-2)$ has a double peak profile and high-velocity emission wings extending to $V_{\rm lsr}=29$ and 53 km s$^{-1}$ (above the noise level) for the blue and red wings, respectively. However, its double peaks have different asymmetry with the $^{12}$CO lines, i.e., the red peak is slightly stronger than the blue one. The offset spectra of the HCN (Figure 2b) have much lower signal-to-noise ratio (mainly due to their shorter integration time). However, they still evidently show double peak profiles, and have similar intensities for the blue and red peaks. Like in the case of the CO lines, the optically thin isotopic lines, i.e., DCN $(3-2)$ and $(4-3)$ are both single-peaked, hence the HCN $(3-2)$ profile should originate from a self absorption effect. Compared to the $^{12}$CO profiles, the HCN spectra may reflect different gas motions, including core expansion or/and rotation~\citep{pav08}. In particular, as the most possible case, a cold and spherically expanding envelope would cause a prominent blueshifted self-absorption towards the center, while at the offset positions, the expansion should be on the plane of the sky, thus show a less blueshift due to the lower radial velocities. This scenario can reasonably explain the observed line profiles, but still needs a further examination with an improved angular resolution and spectral sensitivity. In Figure 2b, we also plotted the velocity-integrated map of HCN $(3-2)$ (discrete gray scales). We measured the average deconvolved radius $\langle r \rangle$ of HCN from its 50 \% contour. As a result, it has $\langle r \rangle=12''$ (and $\eta_{\rm bf}=1$).

Another JCMT line, H$^{13}$CO$^+$ $(3-2)$, has a regular gaussian profile, indicating that it may arise from the dense molecular core and is not evidently affected by the infall or outflow motion. We only have one-point observation for the H$^{13}$CO$^+$ $(3-2)$ at the continuum center, and in calculation for its column density (Section 4.4), we assume $\eta_{\rm bf}=1$.

% more literatures about the DCN production
% 画出C34S和CH3OH延长的方向
% fig 1加上IRAC RGB image

\section{Discussion}
\subsection{The physical properties of the dust core}
As shown in Figure 4, the 1.3 mm dust core does not coincide with any infrared stellar sources besides the extended 4.5 $\micron$ shock emission. This indicates that the stellar emission from the core center is highly obscured by the dust. In the vicinity of the 1.3 mm dust core, a few stellar objects are shown in the IRAC RGB image (also labeled with the asterisks). All these objects are isolated from the 1.3 mm continuum emission, yet the three objects nearest to the continuum peak are likely to be embedded in the 450 $\micron$ emission region. They might either be more evolved young stars in the same star forming region or irrelevant foreground stars. Despite this uncertainty, it is clear that these objects have no significant contribution to the dust continuum or molecular line emissions. We therefore make no further discussion for them.

We can fit the Spectral Energy Distribution (SED) of the dust core from its flux densities at the Spitzer and JCMT/SCUBA wavebands. Assuming a gray body emission and a uniform dust temperature $T_{\rm d}$, the continuum flux density would be~\citep{schn07}

\begin{equation} %equation 1
F_{\nu}=\frac{M_{\rm core}\kappa_{\nu}B_{\nu}(T_{\rm d})}{g D^2}
\end{equation}
where $F_{\nu}$ is the flux density at frequency $\nu$. $M_{\rm core}$ is the total gas-and-dust mass of the core. $g=100$ is commonly adopted gas-and-dust mass ratio. $B_{\nu}(T_{\rm d})$ is the Planck function at temperature $T_{\rm d}$. $D=4.5$ kpc is the source distance. The dust opacity $\kappa_\nu$ is assumed to have a power-law shape, i.e. $\kappa_\nu=\kappa_{\rm 230 GHz}(\nu/{\rm 230 GHz})^{\beta}$, with the reference value $\kappa_{\rm 230 GHz}=0.9$ cm$^2$ g$^{-1}$~\citep{ossen94}. The free parameters in the fit are $M_{\rm core}$, $T_{\rm d}$, and $\beta$. We found that the emissions from 8 $\micron$ to 850 $\micron$ can be best fitted with two temperature components which have $T_{\rm d}=20$ K and 120 K respectively, and $\beta=2.1$. The best-fit SED is shown in Figure 6.

We did not include the IRAC 3.6, 4.5 or 5.8 $\micron$ emissions in our SED model. The 4.5 and 5.8 $\micron$ emissions may largely come from the shocked emission thus are much stronger than the emissions at other two IRAC bands (Table 4). As for the 3.6 $\micron$ emission, if being thermally excited, it may arise from some even hotter component which is much fainter and poorly constrained by our current data. Therefore we also neglected the 3.6 $\micron$ band. With the derived SED, the bolometric luminosity can be estimated using $L_{\rm bol}=4\pi D^2\int F_{\nu}{\rm d}\nu$. As a result we have $L_{\rm bol}=2.3\times10^4$ and $8\times10^2$ $L_{\odot}$ for the cold (20 K) and warm (120 K) components, respectively. Using Equation (1), we can also estimate the total mass of the two temperature components, which turn out to be $1.3\times10^3$ and $1.0\times10^{-3}~M_{\odot}$ for the 20 K and 120 K components, respectively. One can see that both the core mass and luminosity are dominated by the gas-and-dust component which is characterized by $T_{\rm d}=20$ K.

In Equation (1), by replacing the integrated flux density $F_{\nu}$ with the flux density at the continuum peak (0.32 Jy beam$^{-1}$), and then dividing the obtained mass with the beam area and the average molecular mass (1.4 times the molecular mass of H$_2$), we can derive the H$_2$ column density $N({\rm H_2})$ towards the continuum peak. And then, assuming that the core is approximately spherical, we can derive the volume number density using $n({\rm H_2})=N({\rm H_2})/2\langle r\rangle$. The physical parameters of the dust core are presented in Table 4.

By extrapolating the best-fit SED curve, we can get a flux density of 2.9 Jy at $\lambda=1.3$ mm. Compared to this value, the SMA observation has recovered 35\% of the 1.3 mm continuum emission. Adopting $T_{\rm d}=20$ K, we also estimated the physical parameters from the SMA 1.3 mm continuum, which are presented in Table 4. Based on the continuum observations, we suggests that the gas-and-dust core in G8.68 should consist of a dense inner region (characterized by the 1.3 mm emission), and a more extended envelope (traced by the 450 $\micron$ emission).

\subsection{The CH$_3$OH rotational temperature}
The molecular gas temperature can also be estimated from the CH$_3$OH lines using the rotation diagram. The methanol lines all have linewidths of several km s$^{-1}$, with none of them showing abnormally high intensities, therefore the CH$_3$OH lines are unlikely to have maser excitations.

Assuming optically thin, the column density of the upper-level $N_{\rm u}$ can be derived from the integrated line intensity using~\citep{tie05}

\begin{equation}% eq.2
N_{\rm u}=\frac{8\pi k \nu_{\rm ul}^2}{h c^3 A_{\rm ul}}\int T_{\rm b} {\rm d}V /\eta_{\rm bf}
\end{equation}
where $T_{\rm b}$ is the observed brightness temperature. $A_{\rm ul}$ is the Einstein coefficient in s$^{-1}$. $\eta_{\rm bf}$ is the beam filling factor. All other constants take their usual values in SI units. Although for an emission line, $\eta_{\rm bf}$ may vary at different velocities, we approximate it to be a single value as the ratio between the integrated emission region and the beam area, i.e., $\eta_{\rm bf}\simeq\pi\langle r\rangle^2/A_{\rm beam}$ (used when the emission region is smaller than the beam size, otherwise $\eta_{\rm bf}=1$). $\eta_{\rm bf}$ is estimated for each transition and the derived values are presented in Table 5 (Column 11).

Assuming a Local Thermal Equilibrium (LTE, i.e. energy levels are populated according to a Boltzmann distribution characterized by a single temperature), the relation between the total column density $N_{\rm T}$ and $N_{\rm u}$ is

\begin{equation}% eq.3
\frac{N_{\rm u}}{g_{\rm u}}=\frac{N_{\rm T}}{Q(T_{\rm rot})}\exp({-\frac{E_{\rm u}}{k T_{\rm rot}}})
\end{equation}
and its logarithmic form is
\begin{equation}% eq.4
\ln(\frac{N_{\rm u}}{g_{\rm u}})=\ln(\frac{N_{\rm T}}{Q(T_{\rm rot})})-\frac{E_{\rm u}}{k T_{\rm rot}}
\end{equation}
where $g_{\rm u}$ and $E_{\rm u}$ are the degeneracy and the excitation energy of the upper level, respectively, and $Q(T_{\rm rot})$ is the partition function. For CH$_3$OH, a good approximation is $Q(T_{\rm rot})\simeq1.2327\times T_{\rm rot}^{1.5}$ \citep{town55}.

The rotation diagram for the CH$_3$OH lines is shown in Figure 7. A linear least-square fit to the data points results in $T_{\rm rot}=130\pm10$ K and $N_{\rm T}=(5.3\pm0.6)\times10^{15}$ cm$^{-2}$.

In the calculation, in order to correct for the optical depth effect, one should multiply $N_{\rm u}/g_{\rm u}$ with a correction factor $C_{\tau}=\tau/(1-{\rm e}^{-\tau})$ and fit the rotational temperature iteratively. The optical depth is estimated using~\citep[][Equation (3) therein, slightly reformed]{rem04}

\begin{equation} %equation 5
\tau=\frac{c^3\sqrt{4\ln2}}{8\pi\nu^3\sqrt{\pi}\Delta V}N_{\rm u}A_{\rm ul}[\exp(\frac{h\nu}{k T_{\rm rot}})-1]
\end{equation}
Among all the CH$_3$OH lines, the $(8_{-1}-7_{0})$ transition has the highest optical depth ($\tau=0.059$). The other lines are even more optically thin. To take into account the temperature uncertainty, we also estimated the optical depth assuming $T_{\rm rot}=20$ K which is a lower limit as suggested by the SED fitting. At 20 K, the optical depths become $\sim8$ times larger than the values at $T_{\rm rot}=130$ K. The derived optical depths are listed in Table 5, and the column densities and abundances are listed in Table 6.

The rotational temperature of $T_{\rm rot}=130$ K is close to the SED temperature of the warm dust component ($T_{\rm d}=120$ K). Therefore it is possible that the CH$_3$OH emissions are mainly from the region associated with the warm dust. Moreover, since the cold component ($T_{\rm d}=20$ K) is more massive than the warm one for orders of magnitude, the CH$_3$OH may have a severe depletion in the region for the cold dust component. However, it is also possible that the dust and gas are thermally decoupled, thus exhibit different temperatures. The collisional excitations of the molecular gas can be particularly enhanced by the shocks (especially along the outflow direction), thereby showing a high value of $T_{\rm rot}$. The dust temperature $T_{\rm d}$, in comparison, may still be largely dominated by the stellar heating thus has a much lower value.

\subsection{The CO emission and the infall motion}
As shown in Section 3.2.2, both the infall and outflow signatures are detected in the $^{12}$CO $(2-1)$ and $(1-0)$ lines. In this paper we mainly discuss the infall properties based on the $^{12}$CO $(2-1)$. We first make attempt to separate the different components from the observed spectrum, then estimate the infall rate.

Following the procedure of \citet{pur06}, we used a broad gaussian profile to fit the outflow wings (velocity range of $V<32$ and $V>42$ km s$^{-1}$), and then subtracted it from the spectrum. The residual line profile (green line in Figure 8) should mainly represent the emission from the dense molecular core. One can then mask the velocity range possibly affected by the infall motion (34 to 43 km s$^{-1}$), and make a gaussian fit to the spectrum outside this velocity range. The fitted spectrum is speculated to roughly represent the molecular core emission unaffected by the infall signature. However, for the $^{12}$CO lines, due to its large optical depth, we cannot directly apply a Gaussian fit to the spectrum. Instead, one should model the spectrum using the radiation transfer function. In this case, the line profile can be expressed as

% eq.6
\begin{equation}
T_{\rm mb}(V)=[T_{\rm mb,0}-J(T_{\rm CMB})][1-{\rm e}^{-\tau(V)}]
\end{equation}
$J(T)=T_0/[\exp(T_0/T)-1]$ is the Planck-corrected brightness temperature, and $T_0=h\nu/k$. $T_{\rm CMB}=2.7$ K is the temperature of the cosmic background. At the frequency of CO $(2-1)$ (230 GHz), we have $J(T_{\rm CMB})=0.2$ K. Compared to the intensity of the CO emission, the contribution from the cosmic background can be almost neglected.

We also assume the dense molecular core to have a uniform gas distribution along the line of sight, with central velocity $V_0$, velocity dispersion $\sigma$ and peak optical depth $\tau_0$. Then the optical depth is

\begin{equation}% eq.7
\tau(V)=\tau_0\exp[-\frac{(V-V_0)^2}{2\sigma^2}]
\end{equation}
where $\sigma$ is related to the (intrinsic) line width $\Delta V$ by $\sigma=\Delta V/\sqrt{8\ln2}$. In an optically thick case, the line emission could be largely saturated. $\tau_0$ is thus poorly constrained by the observed spectrum. However, it can be estimated from comparison to the CO isotopologues following \citet[][Equation (4) therein]{gard91}. Since the $^{13}$CO is also affected by the infall motion, we used the C$^{18}$O $(2-1)$ instead. Assuming an abundance ratio of $[{\rm ^{12}CO}/{\rm C^{18}O}]=490 $\citep{gard91}, the equation will be

\begin{equation}% eq.8
\frac{T_{\rm mb,0}({\rm ^{12}CO})}{T_{\rm mb,0}({\rm C^{18}O})}=\frac{1-\exp[-\tau_0({\rm ^{12}CO})]}{1-\exp[-\tau_0({\rm C^{18}O})]}=\frac{1-\exp[-\tau_0({\rm ^{12}CO})]}{1-\exp[-\tau_0({\rm ^{12}CO})/490]}
\end{equation}

To fit the line profile, we first take an arbitrary, but reasonable value of $\tau_0$, and then fit the line profile by adjusting the values of $T_{\rm mb,0}$, $V_{\rm 0}$ and $\Delta V$ in Equation (6) and (7). The best-fit $T_{\rm mb,0}$ is then used to estimate $\tau_0$ again using Equation (8). The final best fit can be reached after two or three iterations. Eventually, we have $T_{\rm mb,0}=32$ K, $\Delta V=5.5$ km s$^{-1}$, $V_0=37$ km s$^{-1}$, and $\tau_0({\rm ^{12}CO})=68$. In Figure 8, the best fit spectrum is shown in dashed line. And a sum of dense-core and outflow components is shown in red dot-dashed line. The output spectrum has an apparent line width of 8.5 km s$^{-1}$ which is indeed much broader than the intrinsic $\Delta V$. We fit the $^{12}$CO $(1-0)$ using the same method. All their line parameters are listed in Table 5 (Column 5 to 8).

The infall rate is estimated using \citep{kw07}
\begin{equation}  %eq.9
\dot{M}_{\rm inf}=\frac{4}{3}\pi n({\rm H_2})\mu m_{\rm H} r_{\rm gm}^2 V_{\rm in}
\end{equation}
wherein $r_{\rm gm}$ is geometric mean radius of the core, $n({\rm H_2})$ is the ambient source density, and $V_{\rm in}$ is the typical infall velocity. In calculation we estimated $V_{\rm in}$ from the outflow-subtracted line profile using Equation (9) in \citet{my96}. As a result we have $V_{\rm in}=0.45$ km s$^{-1}$. In addition, we assume that the more diffused gas traced by the 450 $\micron$ emission which has $n({\rm H_2})=0.8\times10^6$ cm$^{-3}$, is collapsing towards the dense inner region characterized by the 1.3 mm continuum (Figure 4), thus we have $r_{\rm gm}=\langle r\rangle_{\rm 1.3 mm}=0.08$ pc. With these assumptions, we derived an infall rate of $\dot{M}_{\rm inf}=7.0\times10^{-4}~M_{\odot}$ yr$^{-1}$.

As seen in Equation (9), the derived infall rate is sensitive to the adoption of $r_{\rm gm}$, and our currently adopted $r_{\rm gm}$ is relatively conservative. Adopting $r_{\rm gm}=\langle r\rangle_{450 \micron}=0.23$ pc, we would have $\dot{M}_{\rm inf}=5\times10^{-3}~M_{\odot}$ yr$^{-1}$. However, we note that such a large-scale estimate may deviate from the small-scale infall rate. To resemble the mass infall onto the central stars, it may be more reasonable to adopt the first value ($r_{\rm gm}=0.08$ pc). With the obtained infall rate, we then estimate the accretion luminosity, using $L_{\rm acc}=GM_*\dot{M}_{\rm inf}/R_*$, and assuming a mass-radius relation of $R_*/R_\odot=(M_*/M_\odot)^{0.8}$. As a result, we have $L_{\rm acc}=(4\pm2)\times10^4~L_{\odot}$. The uncertainty in $L_{\rm acc}$ corresponds to a stellar mass varying between 10 an 100 $M_{\odot}$. It is likely that the bolometric luminosity of the dust core ($2.3\times10^4~L_{\odot}$, see Section 4.1) should have a major energy supply from the accretion process.

Considering the existence of the outflow, there should be a strong interaction between the infall and the outflow. And the interaction may be responsible for the 4.5 $\micron$ shock emission. \citet{chen10} have performed an HCO$^+$ (1-0) survey towards the Extended Green Objects (EGOs), i.e., the massive YSO candidates with the 4.5 $\micron$ shock emissions. They found that nearly one third of the sample (29 out of 69 sources) exhibit a significant blue asymmetry. While in an HCO$^+$ $(1-0)$ survey towards 82 massive YSOs selected from the methanol masers, only 12 sources have infall signatures~\citep{pur06}. Comparing these results, it is likely that the shocks are prone to take place in the YSOs with infall motions. This is theoretically expected, since compared to an interaction between the outflow and quiescent gas, an outflow-infall interaction would more efficiently convert the kinetic energy into heat and radiation.

% more sophisticated model gives a smaller infall rate.
% compare CO (2-1) and (1-0): infall becomes more significant on small scales.

\subsection{The molecular abundances and DCN overabundance}
The total column density $N_{\rm T}({\rm X})$ and abundance $f({\rm X})=N_{\rm T}({\rm X})/N({\rm H_2})$ of the C$^{18}$O, HCN, DCN, H$^{13}$CO$^+$, and C$^{34}$S are calculated from their emission lines at the continuum emission peak using equation (2) and (3). And a correction for the optical depth is done using Equation (4). To derive the HCN and H$^{13}$CO$^+$ abundances, we used the $N({\rm H_2})$ value for the SCUBA 450 $\micron$ continuum (Table 3). While for the SMA lines, we adopted $N({\rm H_2})$ from the 1.3 mm continuum ($0.95\times10^{24}$ cm$^{-2}$). We also note that $N_{\rm T}({\rm H^{13}CO^+})$ may be underestimated due to the beam dilution thus should be regarded as a lower limit. To take into account the temperature uncertainty, we also estimated $N_{\rm T}({\rm X})$ at the lower limit of $T_{\rm rot}=20$ K (suggested by the SED fitting). The $N_{\rm T}({\rm X})$ and $f({\rm X})$ values are shown in Table 6.

In calculation of the HCN, its line profile should be corrected for the self absorption. We modeled its original line profile using the same method for the $^{12}$CO $(2-1)$ (Section 4.2). However, since the abundance ratio between DCN and HCN is much more uncertain than [C$^{18}$O/$^{12}$CO], we cannot use DCN to reliably determine the optical depth of HCN $(3-2)$. Nevertheless, we expect the HCN $(3-2)$ to have a low optical depth due to two reasons. Firstly, since the HCN $(3-2)$ likely traces denser and hotter gas than the $^{12}$CO $(2-1)$, if the HCN $(3-2)$ has a very large optical depth, it should have a comparable intensity with the $^{12}$CO $(2-1)$. Nevertheless, even after the self-absorption correction, the HCN $(3-2)$ is still much weaker than the $^{12}$CO $(2-1)$. Second, with an apparent line width ($\Delta V=6.2$ km s$^{-2}$) being close to $\Delta V$ of the C$^{18}$O and CH$_3$OH lines (as shown in Table 5), the optical-depth broadening should be insignificant for the HCN $(3-2)$. We therefore directly calculate $N_{\rm T}({\rm HCN})$, and then estimate the optical depth using Equation (5). As a result, we found $\tau=0.78$ at $T_{\rm rot}=20$ K and 0.10 at 130 K. This result is consistent with our expectation. However, to more accurately determine the HCN optical depth, one should consider to observe some other isotopologues such as HC$^{15}$N~\citep{hat98}.

From the derived abundances, we can get a relative abundance ratio between DCN and HCN which is [DCN/HCN]$\simeq0.07$. The values derived at the two temperature limits are similar to each other (Table 6). Compared to the cosmic [D/H] ratio \citep[$10^{-5}$,][]{lin98}, the [DCN/HCN] in G8.68 implies a deuteration for orders of magnitudes. The [DCN/HCN] in G8.68 is also much higher that the values detected in hot molecular cores~\citep[$10^{-4}$ to $10^{-3}$,][]{hat98}. However, it is much more comparable to the abundance ratio of [N$_2$D$^+$/N$_2$H$^+$] detected in high-mass YSOs in the infrared dark clouds~\citep{chen11}. Overabundant DCN was previously detected in a number low-mass YSOs~\citep{rob02}, while in high-mass star-forming regions, the DCN is not frequently detected.

The overabundant DCN in G8.68 may originate from a high level of deuterium fractionation in the previous cold pre-stellar phase. In highly deuterated gas (abundant in H$_2$D$^+$, CH$_2$D$^+$, C$_2$HD$^+$ etc.), DCN can be produced via D-H substitution of the HCN, or through more complex reaction pathways~\citep[][Reaction (17) to (21) therein]{alb11}. Finally the DCN molecules would mostly reside on the grain mantles~\citep{hat98,rob02}. During the star formation, the DCN can be released into the gas-phase again. However, once the temperature slightly increases, the gas-phase DCN can be easily destroyed via reactions such as \\ ${\rm H+DCN\rightarrow HCN+D}$~\citep[][Figure 5 therein]{ch92,rob02}. In this sense, to maintain the DCN overabundance in the gas, two conditions may have to be satisfied. First, there should be a high-level deuterium fractionation in the previous dark-cloud phase. Second, in order to compensate the chemical destruction due to the stellar heating, a rapid sublimation for the dust grains should be necessary. Again, the outflow and shocks may have a potential contribution to this process. However, unlike the C$^{34}$S and CH$_3$OH, the DCN emission has a compact and spherical morphology which is not evidently coherent with the outflow. Therefore it is possible that the DCN enhancement is less affected by the shocks and/or more sensitive to the stellar heating. A higher sensitivity and spatial resolution may help better reveal the DCN morphology and determine whether it is associated with the outflow.

As another possibility, the DCN can also be synthesized in the recent gas phase chemistry. \citet{par09} show that the gas-phase reactions may sufficiently account for the enhanced D-H ratio in the molecular gas in the Orion Bar PDR, which has a [DCN/HCN] ratio of $10^{-2}$, comparable with that in G8.68. However, the gas-phase enhancement may have to proceed in an environment with stable lukewarm heating. This condition may hardly be satisfied in regions with rapid evolution of the high-mass stars. Therefore the grain mantle sublimation may still be the major process for the DCN enhancement in G8.68. In the future study, one can compare other chemical products from the grain sublimation and gas-phase chemistry in order to determine the relative importance of these two processes.

% We also note that our HCN-to-H$_2$ abundance ($\sim10^{-11}$) is much lower than the

The C$^{34}$S abundance in G8.68 is much lower than the average C$^{34}$S abundance in the UC H{\sc ii} regions~\citep{olmi99}. $f({\rm C^{18}O})$ is also much smaller than the typical ISM value \citep[$1.7\times10^{-7}$,][]{frer82}. Compared to the ISM abundance, it has a depletion factor of $f_{\rm D}({\rm C^{18}O})=5\pm2$.

\section{Summary}
We have investigated the dust continuum and molecular line emissions towards the high mass YSO G8.68-0.37. We have revealed a dense compact gas-and-dust core in the SMA 1.3 mm continuum emission, and its more extended envelope in the SCUBA 450 $\micron$ emission. At our angular resolution (spatial scale $>0.05$ pc), there is no evident fragmentation structures. We find that an SED with at least two temperature components is necessary to account for the dust continuum emissions from mid-IR to submillimeter wavelengths. The best-fit temperatures for the two components are $T_{\rm d}=20$ K and 120 K. The core mass and luminosity are mainly contributed by the cold component ($T_{\rm d}=20$ K).

Prominent infall signatures and outflow wings are detected in both $^{12}$CO $(1-0)$ and $(2-1)$ lines. We separated the outflow and dense-core components and measured their line parameters. The $^{12}$CO $(2-1)$ yields an infall velocity of 0.45 km s$^{-1}$. Assuming that the extended envelope is collapsing towards the inner dense region, we can derive an infall rate of $7\times10^{-4}$ $M_{\odot}$ year$^{-1}$. It is possible that the 4.5 $\micron$ shock emission is largely enhanced by a strong interaction between the infall and outflow motions. In addition, we also suggest that the infall motion may be important for suppressing the stellar emissions thereby protecting the DCN and other fragile species.

We estimated a rotational temperature of 130 K from the CH$_3$OH lines. We derived the abundances of the molecular species from their spectra, and in particular, we found a high abundance ratio of [DCN/HCN]$=0.07$. The over abundant DCN may originate from a high-level of deuterium fractionation in the previous pre-protostellar phase, as well as the recent grain mantle sublimation and/or gas-phase chemistry. More details in this chemical process are still to be further investigated.

% infall has suppressed the stellar emission
% 没有看到盘啊什么的，可能是演化阶段过早
% stellar heating largely suppressed by the infall motion

\section*{Acknowledgment} We are grateful to the SMA observers and the SMA data archive. We would thank the anonymous reviewer for the detailed, thoughtful comments and suggestions helping us to largely improve the presentation and interpretation. This work is supported by the NSFC grants of No.10733033, 10873019, 10973003, and the NKBRP grandts of 2009CB24901 and 2012CB821800.

% Doctoral Candidate Innovation Research Support Program (kjdb201001-1) from \textit{Science \& Technology Review}.

\clearpage

%table 1
\begin{table*}
%\centering
\begin{minipage}{100mm}
\caption{General information of the single dish observations.\label{tbl1}}
\begin{tabular}{lccc}
\hline\hline
Instrument         &  PMO                     &  CSO                   &  JCMT            \\
%\hline
Obs. date          &  Dec 2009                &  May 2011              &  Aug 2005        \\
Beam size          &  $56''$                  &  $30''$                &  $21''$          \\
$\eta_{\rm mb}$    &  0.62                    &  0.698                 &  0.63            \\
Pointing center    &  R.A.$=~18:06:22.87$     &  R.A.$=~18:06:23.5$    &  R.A.$=~18:06:23.46$  \\
\quad              &  Dec.$=-21:37:20.7$      &  Dec.$=-21:37:10.7$    &  Dec.$=-21:37:09.64$  \\
\hline
\end{tabular}
\end{minipage}
\end{table*}

%table 2
\begin{table*}
%\centering
\begin{minipage}{150mm}
\caption{observational parameters for the molecular lines from the single dishes.\label{tbl1}}
\begin{tabular}{lccccc}
\hline\hline
Transition              &  Instrument   &  Atmosphere   &  Band width   &  $\Delta V_{\rm res}$  &  rms noise   \\
\quad                   &  \quad        &  opacity      &  (MHz)        &  (km s$^{-1}$)         &  per channel (K)    \\
\hline
$^{12}$CO $(2-1)$       &  CSO          &  0.167        &  500          &  0.079                 &  0.2      \\
$^{13}$CO $(2-1)$       &  CSO          &  0.152        &  500          &  0.083                 &  0.2      \\
C$^{18}$O $(2-1)$       &  CSO          &  0.149        &  500          &  0.083                 &  0.2      \\
$^{12}$CO $(1-0)$       &  PMO          &  0.015        &  145          &  0.370                 &  0.1      \\
$^{13}$CO $(1-0)$       &  PMO          &  0.015        &  43           &  0.115                 &  0.1      \\
C$^{18}$O $(1-0)$       &  PMO          &  0.015        &  43           &  0.115                 &  0.2      \\
HCN $(3-2)$             &  JCMT         &  0.111        &  160          &  0.088                 &  0.5      \\
H$^{13}$CO$^+$ $(3-2)$  &  JCMT         &  0.067        &  160          &  0.090                 &  0.3      \\
\hline
\end{tabular}
\end{minipage}
\end{table*}

%table 3
\begin{table*}
%\centering
\begin{minipage}{170mm}
\caption{Observational parameters of the SMA.\label{tbl3}}
\begin{tabular}{lccllll}
\hline\hline
Epoche       &  Frequency bands (GHz)         &  Bandpass      &  Flux         &  Phase \& synthesized   & beam size   & rms noise  \\
\quad        &  LSB, USB                      &  Calibrator    &  Calibrator   &  Calibrator             & (arcsec)    & per channel (K)$^a$ \\
\hline
2007         &  (279.4,281.4), (289.4,291.4)  &  3c273         &  Neptune      &  1733-130,1911-201      &  $7.0\times5.8$     & 0.110   \\
2008         &  (219.5,221.5), (229.5,231.5)  &  3c454.3       &  Neptune      &  1733,1911              &  $4.8\times3.6$     & 0.017  \\
2009         &  (217.5,219.5), (227.5,229.5)  &  3c273         &  Uranus       &  1733,1911              &  $6.8\times3.6$     & 0.150   \\
\hline \\
\end{tabular} \\
$a.${ For the unit conversion, 1 K=0.367, 1.43 and 1.04 Jy beam$^{-1}$ for the data in 2007, 2008, and 2009 respectively.}
\end{minipage}
\end{table*}

%table 4
\begin{table*}
%\centering
\begin{minipage}{80mm}
\caption{Physical parameters of the dust core.\label{tbl3}}
\begin{tabular}{ccl}
\hline\hline
Parameter             &  Value                &   Unit   \\
\hline
$F(3.6 \micron)$      &  $13\pm0.1$           &   mJy    \\
$F(4.5 \micron)$      &  $78\pm2$             &   ...    \\
$F(5.8 \micron)$      &  $68\pm2$             &   ...    \\
$F(8.0 \micron)$      &  $10\pm1$             &   ...    \\
$F(24  \micron)$      &  $657\pm40$           &   ...    \\
$F(70 \micron)$       &  $196\pm15$           &   Jy     \\
$F(450 \micron)$      &  $144\pm15$           &   ...    \\
$F(850 \micron)$      &  $15\pm4$             &   ...    \\
$F(1.3 {\rm mm})^a$   &  $0.65\pm0.03$        &   ...    \\
$F(1.3 {\rm mm})^b$   &  $2.9$                &   ...    \\
\hline
\quad                 &  $(450~\micron$ / 1.3 mm)$^c$   &   \quad                 \\
$\langle r\rangle $   &  $0.23\pm0.05$ / $0.08\pm0.02$  &   pc                    \\
$M_{\rm core}$        &  $1.5\pm0.2$ / $0.30\pm0.01$    &   $10^3~M_{\odot}$      \\
$N({\rm H_2})$        &  $1.2\pm0.2$ / $0.95\pm0.03$    &   $10^{24}$ cm$^{-2}$   \\
$n({\rm H_2})$        &  $0.8\pm0.1$ / $~3.8\pm0.3~$    &   $10^6$ cm$^{-3}$      \\
\hline
\end{tabular} \\
{\small
Note. To measure the integrated flux density of the dust core, we use the 4 $\sigma$ emission level as the integration area (aperture for the photometry). As an exception, we use the 4 $\sigma$ level of the 4.5 $\micron$ emission as the area for all four IRAC bands, which roughly equals to the emission region with green color in Figure 4. The nearby point sources carefully excluded from this aperture.\\
$a.${ The flux density of the SMA continuum observation.} \\
$b.${ The flux density extrapolated from the SED fitting (Figure 6).} \\
$c.${ For the last 4 parameters, the first and second values are derived from the 450 $\micron$ and 1.3 mm continuum data, respectively.} \\
}
\end{minipage}
\end{table*}

%table 5
\begin{table*}
%\centering
\begin{minipage}{160mm}
\caption{Observed parameters of the molecular lines.\label{tbl4}}
\begin{tabular}{llccccccccc}
\hline\hline
Molecule              &  Transition             &  Rest frequency  &  $E_{\rm u}$   &  $V_{\rm LSR}$   &  $T_{\rm b,peak}$   &  $\Delta V_{\rm FWHM}$  &  $\int T_{\rm b}{\rm d}V$ &  $\tau^a$        &  $\langle r\rangle ^b$  &  $\eta_{\rm bf}^c$ \\
\quad                 &  \quad                  &  (GHz)           &  (K)           &  (km s$^{-1}$)   &  (Kelvin)           &  (km s$^{-1}$)          &  (K km s$^{-1}$)          &  \quad           &  (arcsec)               &  \quad           \\
(1)                   &  (2)                    &  (3)             &  (4)           &  (5)             &  (6)                &  (7)                    &  (9)                      &  (8)             &  (10)                   &  (11)            \\
\hline
$^{12}$CO(core)       &  $1-0$                  &  115.27120       &  5.5           &  37.0            &  26.0(1.5)          &  5.0(0.5)               &  134(15)                  &  75              &  --                     &  --      \\
$^{12}$CO(outflow)    &  $1-0$                  &  ...             &  ...           &  37.3            &  4.0(1.5)           &  9.2(1.0)               &  37(5)                    &  0.15            &  --                     &  --      \\
$^{12}$CO(core)       &  $2-1$                  &  230.53800       &  17            &  37.0            &  32.0(0.5)          &  5.5(0.5)               &  240(30)                  &  68              &  --                     &  --      \\
$^{12}$CO(outflow)    &  $2-1$                  &  ...             &  ...           &  37.0            &  5.0(0.5)           &  9.0(0.5)               &  45(6)                    &  0.15            &  --                     &  --      \\
HCN(core)             &  $3-2$                  &  265.88643       &  26            &  38.5            &  12.0(0.2)          &  6.2(0.5)               &  73(9)                    &  0.78/0.10       &  12                     &  1       \\
HCN(outflow)          &  $3-2$                  &  ...             &  ...           &  39.5            &  0.8(0.2)           &  15.0(1.0)              &  12(5)                    &  0.053/0.007     &  --                     &  --      \\
$^{13}$CO             &  $1-0$                  &  110.20135       &  5.3           &  35.0            &  12.4(0.5)          &  6.8(0.1)               &  85(10)                   &  0.85            &  --                     &  --      \\
$^{13}$CO             &  $2-1$                  &  220.39968       &  16            &  35.0            &  8.5(0.5)           &  7.5(0.5)               &  60(15)                   &  0.70            &  --                     &  --      \\
C$^{18}$O             &  $1-0$                  &  109.78217       &  5.3           &  37.0            &  4.5(0.5)           &  5.6(0.4)               &  23(4)                    &  0.17            &  --                     &  --      \\
C$^{18}$O(CSO)        &  $2-1$                  &  219.56036       &  16            &  38.0            &  6.0(0.3)           &  5.5(0.4)               &  34(3)                    &  0.140           &  --                     &  --      \\
C$^{18}$O(SMA)        &  $2-1$                  &  219.56036       &  16            &  38.0            &  5.2(0.02)          &  4.8(0.3)               &  30(3)                    &  0.423/0.065     &  3.4                    &  1       \\
CH$_3$OH              &  $8_0-7_1$              &  220.07849       &  97            &  37.5            &  1.6(0.22)          &  5.0(1.3)               &  8(2)                     &  0.209/0.028     &  3.1                    &  1       \\
CH$_3$OH              &  $8_{-1}-7_0$           &  229.75880       &  89            &  37.0            &  3.42(0.15)         &  6.0(0.3)               &  18.0(1.5)                &  0.453/0.059     &  2.8                    &  1       \\
CH$_3$OH              &  $3_{-2}-4_{-1}$        &  230.02706       &  39            &  38.5            &  0.9(0.03)          &  6.0(0.4)               &  5.8(0.5)                 &  0.036/0.004     &  3.2                    &  1       \\
CH$_3$OH              &  $10_2-9_3$             &  231.28110       &  165           &  39.0            &  0.53(0.06)         &  6.0(0.9)               &  3.4(0.5)                 &  0.087/0.011     &  2.2                    &  0.95    \\
CH$_3$OH              &  $9_{-1}-8_0$           &  278.30451       &  110           &  38.0            &  1.39(0.06)         &  5.4(0.4)               &  7.2(0.8)                 &  0.245/0.031     &  2.9                    &  0.89    \\
CH$_3$OH              &  $2_{-2}-3_{-1}$        &  278.34226       &  33            &  39.0            &  0.24(0.02)         &  5.5(0.4)               &  1.5(0.2)                 &  0.045/0.006     &  3.2                    &  1       \\
CH$_3$OH              &  $21_{-2}-20_{-3}$      &  278.48023       &  563           &  40.0            &  0.09(0.02)         &  4.0(0.5)               &  0.4(0.05)                &  0.024/0.002     &  2.2                    &  0.48    \\
CH$_3$OH              &  $15_{7,8}-16_{6,11}$   &  288.07677       &  523           &  37.5            &  0.24(0.02)         &  5.5(0.3)               &  1.2(0.2)                 &  0.102/0.012     &  2.3                    &  0.52    \\
CH$_3$OH              &  $14_4-15_3$            &  278.59906       &  340           &  39.5            &  0.29(0.02)         &  6.3(1.3)               &  1.8(0.2)                 &  0.073/0.009     &  2.5                    &  0.62    \\
CH$_3$OH              &  $11_2-10_3$            &  279.35191       &  191           &  39.5            &  0.28(0.02)         &  6.5(1.9)               &  2.3(0.3)                 &  0.081/0.009     &  2.8                    &  0.78    \\
CH$_3$OH              &  $4_3-5_2$              &  288.70557       &  71            &  39.5            &  0.28(0.02)         &  6.5(1.5)               &  1.8(0.3)                 &  0.068/0.008     &  2.8                    &  0.77    \\
DCN                   &  $3-2$                  &  217.23854       &  81            &  39.0            &  2.1(0.4)           &  3.5(0.2)               &  7.5(1)                   &  0.132/0.018     &  3.1                    &  0.92    \\
DCN                   &  $4-3$                  &  289.64492       &  35            &  39.0            &  1.0(0.04)          &  3.5(0.2)               &  5.0(0.4)                 &  0.072/0.009     &  3.3                    &  1       \\
C$^{34}$S             &  $6-5$                  &  289.20907       &  38            &  39.0            &  1.4(0.02)          &  4.5(0.5)               &  6.5(0.7)                 &  0.123/0.016     &  3.3                    &  1       \\
H$^{13}$CO$^+$        &  $3-2$                  &  260.25534       &  25            &  38.0            &  4.7(0.3)           &  4.4(0.3)               &  20(2)                    &  0.260/0.033     &  --                     &  --      \\
\hline \\
\end{tabular}

{\small
Note. The $^{12}$CO, $^{13}$CO lines, and C$^{18}$O lines are from the PMO and CSO observations (see Table 2).
The C$^{18}$O $(2-1)$ line from the SMA observation is also presented. The HCN and H$^{13}$CO$^+$ lines are observed with the JCMT. For the double-peaked lines, including HCN $(3-2)$, $^{12}$CO $(2-1)$ and $(1-0)$, and two CH$_3$OH lines ($E_{\rm u}=191$ K and $E_{\rm u}=523$ K), the parameters are measured from the fitted spectra, while for the single-peak lines, we directly measured the observed spectra. \\
$a.${ The optical depth at the line peak. For the transitions with two values, the first and second one are the results for $T_{\rm rot}=20$ K and 114 K, respectively (see Section 4.4). While for the CO $(1-0)$ and $(2-1)$, the optical depths are calculated from comparing their isotopic lines (Section 4.3).}  \\
$b.${ The effective radius of the emission region, measured from the deconvolved average radius of the 10\% contour region (1/2 times the value). An exception is the HCN $(3-2)$, for which we directly measured the 50 \% contour. The average uncertainty level is $\sim2$ arcsec. For G8.68 at $D=4.5$ kpc, 1 arcsec$=0.02$ pc.} \\
$c.${ The beam filling factor, calculated from the ratio between the area of the deconvolved emission region ($\pi \langle r\rangle ^2$) and the beam size.}
}
\end{minipage}
\end{table*}

%table 6
\begin{table*}
\centering
\begin{minipage}{130mm}
\caption{Collum density and abundance of the molecular species.\label{tbl2}}
\begin{tabular}{lllll}
\hline\hline
\quad                   &  \multicolumn{2}{c}{$T_{\rm rot}=20$ K$^a$}                                    &  \multicolumn{2}{c}{$T_{\rm rot}=130$ K}     \\
Molecule                &  \multicolumn{2}{l}{-----------------------------------------------------}     &  \multicolumn{2}{l}{-----------------------------------------------------}    \\
(X)                     &  $N_{\rm T}({\rm X})$ (cm$^{-2}$)       &  $f({\rm X})$                        &  $N_{\rm T}({\rm X})$ (cm$^{-2}$)        &  \quad$f({\rm X})$                  \\
\hline
C$^{18}$O               &  $(2.2\pm0.3)\times10^{16}$             &  $(2.3\pm0.3)\times10^{-8}$           &  $(5.4\pm0.6)\times10^{16}$             &  $(5.6\pm0.7)\times10^{-8}$      \\
CH$_3$OH$^b$            &  $(1.8\pm0.2)\times10^{15}$             &  $(2.0\pm0.2)\times10^{-9}$           &  $(5.3\pm0.6)\times10^{15}$             &  $(5.8\pm0.6)\times10^{-9}$      \\
C$^{34}$S               &  $(2.3\pm0.3)\times10^{13}$             &  $(2.5\pm0.4)\times10^{-11}$          &  $(2.6\pm0.3)\times10^{13}$             &  $(2.9\pm0.4)\times10^{-11}$     \\
H$^{13}$CO$^+$          &  $(9.5\pm1.0)\times10^{12}$             &  $(7.9\pm0.7)\times10^{-12}$          &  $(1.7\pm0.2)\times10^{13}$             &  $(1.4\pm0.1)\times10^{-11}$     \\
HCN                     &  $(5.6\pm0.5)\times10^{14}$             &  $(4.6\pm0.3)\times10^{-10}$          &  $(1.1\pm0.1)\times10^{15}$             &  $(8.9\pm0.8)\times10^{-10}$     \\
DCN                     &  $(3.1\pm0.1)\times10^{13}$             &  $(3.4\pm0.2)\times10^{-11}$          &  $(5.6\pm0.3)\times10^{13}$             &  $(6.2\pm0.5)\times10^{-11}$     \\
${\rm [DCN/HCN]}^c$   &  --                                     &  $0.07\pm0.01$                        &  --                                     &  $0.07\pm0.01$                   \\
\hline \\
\end{tabular}

{\small
$a.${ A lower limit as suggested by the dust continuum SED.} \\
$b.${ At $T_{\rm rot}=20$ K, $N_{\rm T}({\rm CH_3OH})$ is derived from $3_{-2}-4_{-1}$ line.} \\
$c.${ Abundance ratio between DCN and HCN.}
}
\end{minipage}
\end{table*}
\clearpage

\begin{figure}
\centering
\includegraphics[angle=0,width=0.4\textwidth]{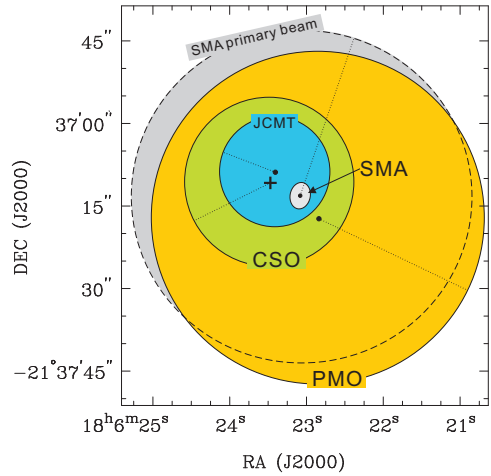} \\
\caption{\small The beam size and pointing center of each instrument. The cross symbol marks the center of the 1.3 mm dust core (Figure 4) which is coincident with the CSO pointing center. The white ellipse is the synthesized beam of the SMA in the 2008 observation, and the gray filled circle is the primary beam. The JCMT beam size is for the frequency of 289 GHz.}
\end{figure}

\begin{figure}
\centering
\includegraphics[angle=0,width=0.8\textwidth]{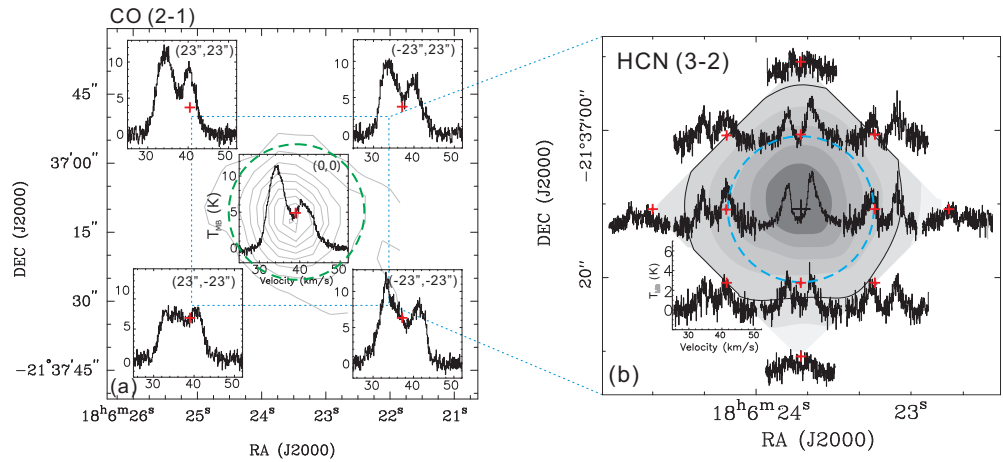} \\
\caption{\small (a) Grid spectra of $^{12}$CO $(2-1)$ observed from the CSO. The red cross labels the pointing center of each spectrum. The gray contours are the SCUBA 450 $\micron$ emission (specified in Figure 4). The green dashed line represents the beam size. (b) Grid spectra of HCN $(3-2)$ observed from the JCMT. The red cross labels the pointing center of the each spectrum. The intensity map (gray scales) is made from interpolating the line intensity at each point. The integration for the spectra is from 25 to 50 km s$^{-1}$. The gray-scale levels are from 30 \% to 90 \% of the maximum intensity (46.8 K km s$^{-1}$). The thick contour is the 50 \% level. The blue dashed circle is the beam size. }
\end{figure}

\begin{figure}
\centering
\includegraphics[angle=0,width=0.8\textwidth]{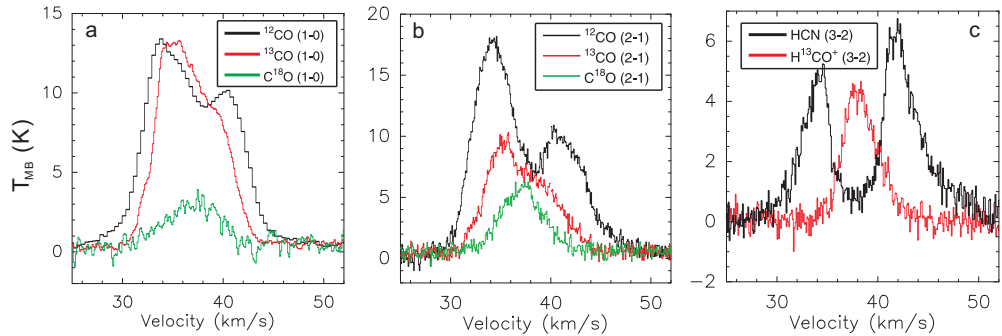} \\
\caption{\small Molecular lines observed from the PMO, CSO, and JCMT, which are shown in left, middle, and right panels, respectively. The observing centers and the beam size of each telescope are shown in Figure 1. }
\end{figure}

\begin{figure}
\centering
\includegraphics[angle=0,width=0.4\textwidth]{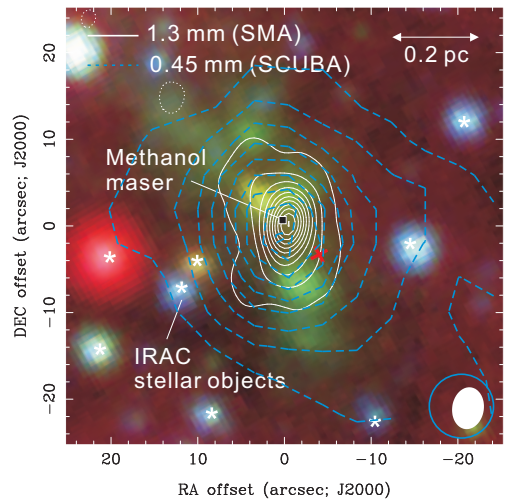} \\
\caption{\small Continuum emissions detected towards G8.68 from infrared to millimeter wavelengths. The image is centered at the emission peak of the 1.3 mm continuum, the coordinates of which are RA.(J2000)=18$^{\rm h}$06$^{\rm m}$23.5$^{\rm s}$ and Decl.(J2000)=$-21^{\circ}37'10.7''$. The white contours are the 1.3 mm emission observed from the SMA. The contour levels are -4, 4, 14, 24... 104 $\sigma$ (0.003 Jy beam$^{-1}$). The -4 $\sigma$ contour is due to the insufficient (u,v) coverage and is plotted in dotted line. The square denotes the strongest CH$_3$OH maser~\citep{walsh98}. The dashed contours are the JCMT/SCUBA 450 $\micron$ emission. The levels are 4, 8, 12... 36 $\sigma$ (1.2 Jy beam$^{-1}$). The IRAC 3.6 (blue), 4.5 (green) and 8.0 (red) $\micron$ images are shown together in RGB colors (also seen in Figure 1 and 2 in L11). The red cross labels the SMA phase center. The synthesized beam of the SMA (white ellipse) and SCUBA (blue circle) beam are plotted in the right corner.}
\end{figure}

\begin{figure}
\centering
\includegraphics[angle=0,width=0.9\textwidth]{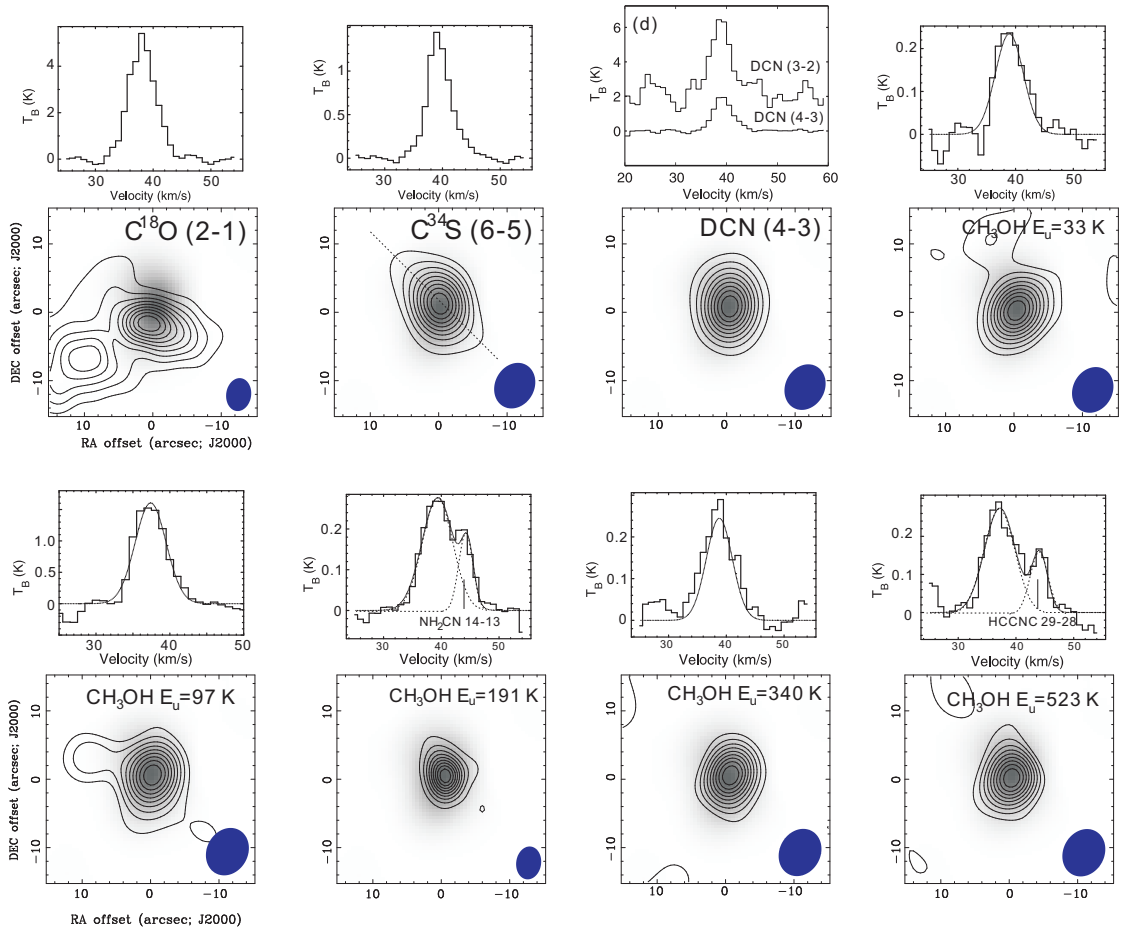} \\
\caption{\small Molecular lines and integrated images observed from the SMA. For each transition, the contours are 10, 20... 90 percent of the peak intensity. The integration range is $(37,~43)$ km s$^{-1}$ for all the transitions except the two blended CH$_3$OH lines. For these two lines the integration range is $(37,40)$ km s$^{-1}$ as to eliminate the contamination. The dashed line in the C$^{34}$S labels the orientation of the outflow in L11. The DCN $(3-2)$ line is shifted above the zero level for 2 K. The negative contours due to the missing flux are omitted to more clearly show the emission features. The gray-scale image in each panel is the SMA 1.3 mm continuum.}
\end{figure}

\begin{figure}
\centering
\includegraphics[angle=0,width=0.5\textwidth]{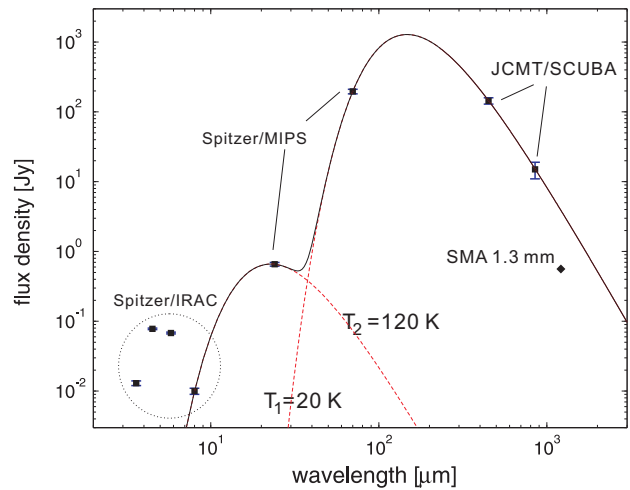} \\
\caption{\small The spectra energy distribution of the dust core. The black squares with error bars are the data points (the SMA 1.3 mm flux density is marked with the diamond). The black line is the fitted SED curve. The red dashed lines are the SEDs of the two temperature components, with $T_{\rm d}=20$ K and 120 K, respectively.}
\end{figure}

\begin{figure}
\centering
\includegraphics[angle=0,width=0.5\textwidth]{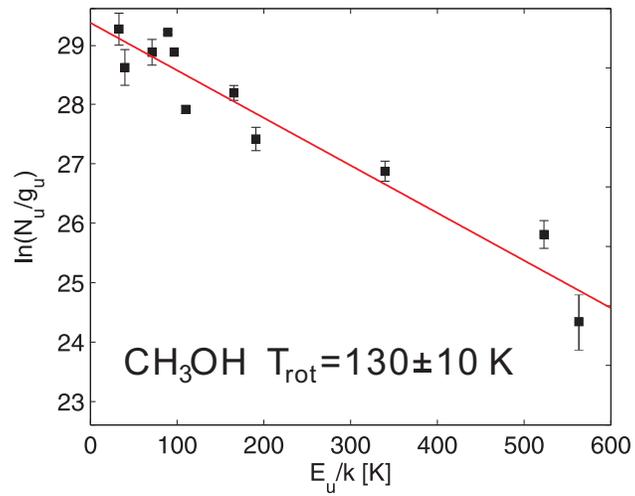} \\
\caption{\small The rotation diagram of the CH$_3$OH lines. The black squares with error bars are the data points. The red line is the least-square fit.}
\end{figure}

\begin{figure}
\centering
\includegraphics[angle=0,width=0.5\textwidth]{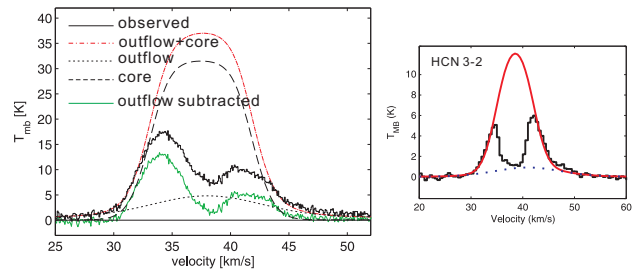} \\
\caption{\small The two-component fit to the $^{12}$CO $(2-1)$ spectrum towards the center. The meaning of each line type is labeled in the legend. }
\end{figure}

\end{document}